\documentclass[twocolumn,aps,showpacs,superscriptaddress]{revtex4}

\usepackage{amssymb}
\usepackage{amsfonts}
\usepackage{amsmath}
\usepackage{graphicx}

\usepackage{bm}
\usepackage{multirow}

\usepackage{graphicx}

\newcommand{\vect}[1]{\boldsymbol{#1}}

\newcommand{\colvect}[2] {\begin{pmatrix} #1 \\ #2\end{pmatrix}}

\begin{document}

\title{Hierarchy of gaps and magnetic minibands in graphene \\ 
in the presence of the Abrikosov vortex lattice}

\author{Xi Chen}
\affiliation{National Graphene Institute, University of Manchester, Booth Street East, Manchester, M13 9PL, UK}
\author{Vladimir I. Fal'ko}
\affiliation{National Graphene Institute, University of Manchester, Booth Street East, Manchester, M13 9PL, UK}
\date{\today}

\begin{abstract}
We determine bands and gaps in graphene subjected to the magnetic field of Abrikosov lattice of vortices in the underlying superconducting film.  
The spectrum features one non-dispersive magnetic miniband at zero energy, separated by the largest gaps in the miniband spectrum from a pair of 
minibands resembling slightly broadened first Landau level in graphene, suggesting the persistence of $\nu = \pm 2$ and $\pm 6$ quantum Hall effect states. 
Also, we identify occasional merging point of magnetic minibands with a Dirac-type dispersion at the miniband edges. 
\end{abstract}
\maketitle


Studies of superlattices in two-dimensional (2D) electron systems have, recently, been boosted by the development of van der Waals heterostructures of graphene with hexagonal boron nitride (hBN). In such systems the superlattice effects, observed in STM spectra \cite{G-hBN-STM1,G-hBN-STM2,G-hBN-STM3}, magneto-transport chracteristics \cite{G-hBN-Tr1,G-hBN-Tr2,G-hBN-Tr3} and quantum capacitance \cite{G-hBN-C}, are produced by a periodic moire pattern, with the period $a$ determined by slight incommensurability and misalignment between graphene and hBN crystals \cite{G-hBN-STM1,Kindermann,Wallbank} and reflect the formation of superlattice minibands for graphene's Dirac electrons \cite{Wallbank,G-hBN-Tr1,G-hBN-C}. To a large extent, the possibility to observe the superlattice effects in graphene-hBN heterostructures owes to the high mobility of electrons in such systems, where graphene is encapsulated between hBN sheets both protecting from contamination and permitting to vary electrons' density over a broad range using electrostatic gates. When subjected to a strong external magnetic field, the superlattice leads to the formation of a 'Hofstadter butterfly', a sparse spectrum of minibands \cite{Zak,Brown,chen_prb_2014} formed at magnetic field values corresponding to the magnetic flux, $\Phi = \frac{p}{q} \phi_0$ (through the area ${\cal S}=\sqrt{3}a^2/2$ of the superlattice unit cell) commensurate with the flux quantum,  $\phi_0 = h/e$.

\begin{figure}[htbp]
\centering
\includegraphics[width=0.44 \textwidth]{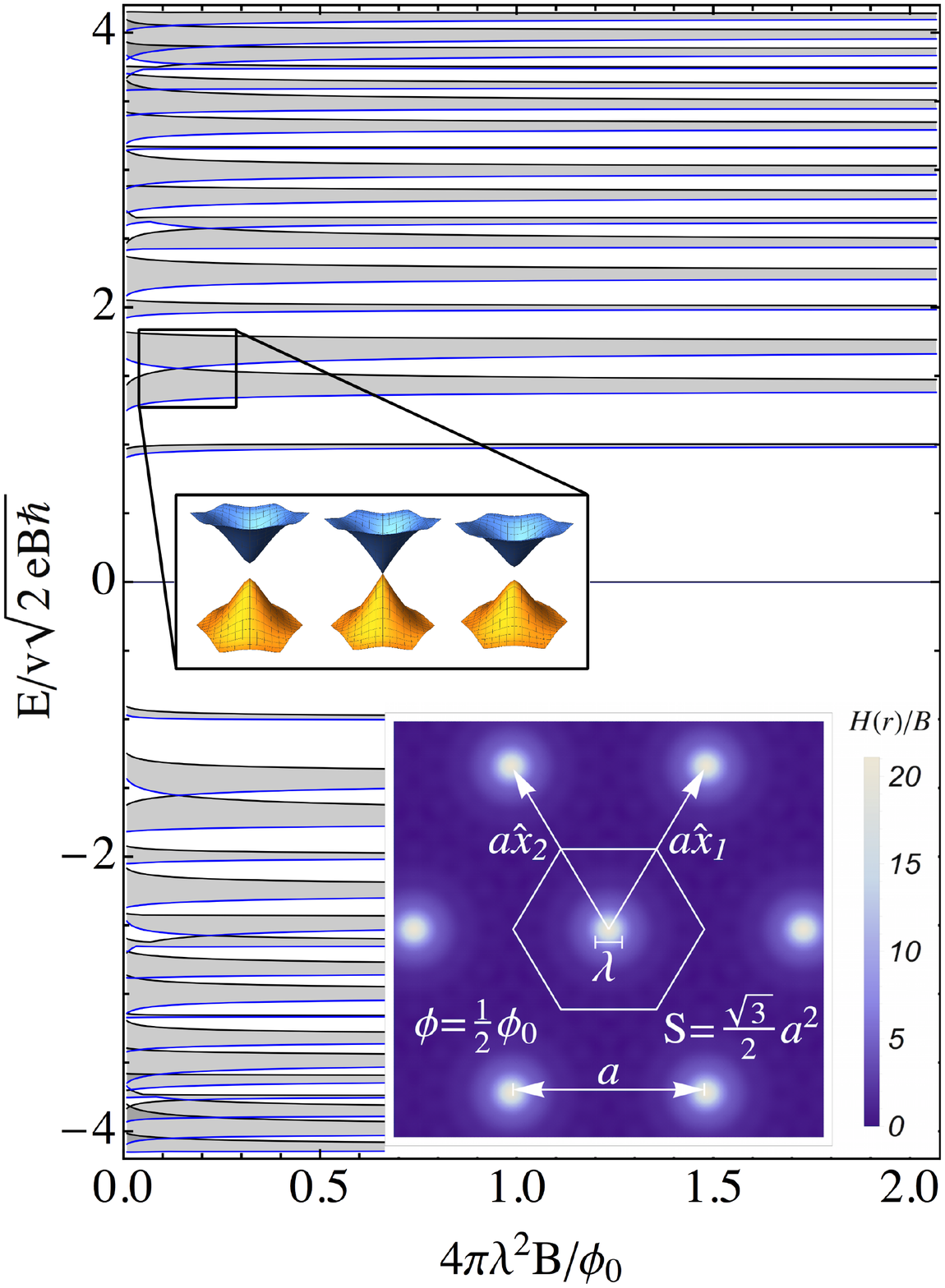} 
\caption{Spectrum of Dirac electrons in graphene in a magnetic field of Abrikosov vortex lattice, with one degenerate band at $E=0$.
Energy is scaled as $v\sqrt{2\hbar e B} \equiv \frac{2}{3^{1/4}}v \sqrt{\pi}/a$, and magnetic field as $4\pi \lambda^2 B/\phi_0$.
Inset: 2nd and 3rd minibands dispersion over the folded magnetic Brillouin minizone near their touching condition.}
\label{fig:VortexSLBandWidth}
\end {figure}

Here, we consider a magnetic superlattice \cite{Vortices1,Vortices2,Vortices3,Vortices4,Vortices5,Vortices6} that can be realised in a ballistic hBN-graphene-hBN stack by placing it over a high-$H_{c2}$ superconductor film ({\it e.g.}, Nb, W, or MoRe alloy). In such a system, where no alignment control of graphene and hBN lattices is required, long-range periodic structure is caused by the Abrikosov lattice of vortices \cite{Abrikosov,Ketterson_book_1999} formed in a superconductor subjected to an external magnetic field $H<H_{c2}$, sketched in the inset in Fig. \ref{fig:VortexSLBandWidth}. In contrast to the earlier theories developed for spatially alternating magnetic fields with a zero average \cite{Taillefumier_PhysRevB_2008,Snyman_PhysRevB_2009,Guinea,Kamfor_PhysRevB_2011,Taillefumier_PhysRevB_2011,Lin}, the Abrikosov lattice produces magnetic induction with spatial average, $B=\phi_0/(\sqrt{3}a^2)$, linked to the magnetic lattice period, $a$. As each vortex carries the flux $h/2e$, the vortex lattice realises the simplest fundamental fraction, $\frac{p}{q} = \frac{1}{2}$, in the Brown-Zak commensurability condition for magnetic field flux in a 2D periodic system \cite{Zak,Brown}. 

Figure \ref{fig:VortexSLBandWidth} shows the hierarchy of bands and gaps in the corresponding spectrum of Dirac electrons calculated in this work and plotted as a function of $4\pi \lambda^2 B/\phi_0$. The latter parameter characterises the the ratio between the lattice period $a=\sqrt{\phi_0/\sqrt{3}B}$ and the penetration depth $\lambda$ in a superconductor. The spectrum in Fig.\ref{fig:VortexSLBandWidth} features one degenerate magnetic miniband which precisely coinsides with the zero-energy $m=0$ Landau level (LL) peculiar for Dirac electrons, and two other low-energy bands which resemble slightly broadened $m=\pm 1$ LLs. At the same time, the higher-energy minibands (traceable at  $4\pi \lambda^2 B/\phi_0 \ge 1$ to LLs with $|m| \ge 1$ at $E= \frac{m}{|m|} v\sqrt{2|m|\hbar e B}$, $m \in  {\mathbb Z}$) are strongly broadened, and they overlap on the energy scale when $a \gg \lambda$, forming Dirac-type features at conjoint miniband edges. 


Distribution of magnetic field of Abrikosov's lattice (for isotropic superconductors, 
the vortex lattice is hexagonal) is given by \cite{Abrikosov,Ketterson_book_1999} :
$$
H({\bf r}) = \sum_{{\bf R}_i} H_v({\bf r}-{\bf R}_i), \quad  
H_v(\vect {r})= \frac{1}{2 \pi} \frac{\phi_0}{2 \lambda^2} K_0(\frac{r}{\lambda}),
$$
where $\vect{R}_i$ picks location of each individual vortex with the field profile given by the modified Bessel function of imaginary argument, with  $K_0(x \ll 1) \approx \ln \frac{1}{x}$ and $K_0(x\gg 1) \approx \sqrt{\frac{\pi}{2x}} e^{-x}$. For convenience, we use Fourier representation for the periodic field of vortex lattice \cite{Ketterson_book_1999} and the corresponding vector potential, 
\begin{align}
\label{eq:LondonEqPeriodicSolution}
&H(\vect{r})
= B\left(1 + \sum_{\vect{g}_{n_1n_2}} \frac{1}{1+ \lambda^2 g_{n_1n_2}^2} e^{-i \vect{g}_{n_1n_2} \cdot \vect{r}}\right),  \nonumber \\
&\vect{A}=\bar{\vect{A}} -\frac{\hbar}{e}\frac{\sqrt{3}}{8 \pi}\sum_{n_1n_2} \frac{\hat{\vect l}_z \times \nabla e^{-i \vect{g}_{n_1n_2} \cdot \vect{r}} }{\omega_{n_1n_2}^2 (1 + \alpha^{\text{-}2} \omega_{n_1n_2}^2)}, \\
&(\vect{\nabla} \times \bar{\vect{A}} )_z = B \equiv \frac{h/e}{\sqrt{3} a^2},   
\qquad    \alpha \equiv \frac{\sqrt{3}a}{4 \pi \lambda},\nonumber
\end{align}
and a non-orthogonal coordinate system adjusted to the hexagonal symmetry of the vortex superlattice, 
\begin{align*}
& \vect{g}_{n_1n_2}= g \hat{\vect z} \times  (n_1 \hat{\vect{x}}_1 + n_2 \hat{\vect{x}}_2), \\
& \omega_{n_1n_2} =\sqrt{n_1^2+n_2^2+n_1n_2}, \quad g =4 \pi/(\sqrt{3} a ) \\
&\hat{\vect{x}}_1=\frac{1}{2}\hat{\vect{x}}+\frac{\sqrt{3}}{2}\hat{\vect{y}}, \quad 
\hat{\vect{x}}_2=-\frac{1}{2}\hat{\vect{x}}+\frac{\sqrt{3}}{2}\hat{\vect{y}},
\end{align*}

The spectrum of electrons in K and K' valleys of graphene is determined by the Hamiltonian 
\begin{align}
\label{eq:MainHofPeriodicVSL}
{\cal H} 
&= v \vect{\sigma} \cdot (\vect{p} -e \bar{\vect{A}}) \nonumber \\
&+ \hbar v \vect{\sigma} \cdot
\frac{\sqrt{3}}{8 \pi}\sum_{n_1n_2} \frac{\hat{\vect l}_z \times \nabla e^{-i \vect{g}_{n_1n_2} \cdot \vect{r}} }{\omega_{n_1n_2}^2 (1 + \alpha^{\text{-}2} \omega_{n_1n_2}^2)},
\end{align}
where $\bar{\vect A} =h x_1(-\hat{\vect{x}}_1 +2\hat{\vect{x}}_2)/(3 a^2 e)$, $x_1 = x + \frac{1}{\sqrt{3}}y$, $x_2 = -x + \frac{1}{\sqrt{3}}y$, and $v \approx 10^6$cm/s is Dirac velocity in graphene. This Hamiltonian acts in the space of two-component wave functions describing electrons' amplitudes on A and B sublattices of the honeycomb lattice of carbons, with the basis choice $[\Psi(A),\Psi(B)]$ in valley K and $[\Psi(B),-\Psi(A)]$ in valley K': this choice provides us with the same form of the  Hamiltonian in both valleys.

To find the magnetic miniband spectrum corresponding to Hamiltonian (\ref{eq:MainHofPeriodicVSL}), we use \cite{chen_prb_2014}  the basis of Bloch states (with $s=-{\cal N}/2, \cdots ,{\cal N}/2$, and $t=0,1$),
\begin{align}
& |_{t}^{m}(\vect k)\rangle  = \frac{1}{\sqrt {\cal N}}\sum_{s} e^{-i 2 s k_1 a} \psi_m^{ k_2+\frac{\sqrt3}{4}g(2s+t)}, \nonumber \\
& k_1=\frac12 k_x +\frac{\sqrt{3}}{2}k_y, \quad k_2=-\frac12 k_x +\frac{\sqrt{3}}{2}k_y,  
 \label{eq:bloch_wf} 
\end{align}
built of LL states $\psi_{m}$, $E_m
=  \frac{m}{|m|} \frac{\hbar v}{a}\sqrt{\frac{\pi}{\sqrt{3}}|m|}$, ($m \in  {\mathbb Z}$) of Dirac electrons in a homogeneous magnetic field $B$ \cite{FootnoteDirac}: 
\begin{align}
\label{eq:H_LLgapless}
&\psi_{0}^{k_2}\!=\! \frac{e^{ik_2 x_2}}{ \sqrt L}\!\colvect{\varphi_0}{0}\! ,\;
\psi_{m \ne 0}^{k_2}\!=\!  \frac{e^{ik_2 x_2}}{\sqrt{2L} }\!\colvect{\varphi _{|m|}}{\frac{m}{|m|}e^{i\frac{\pi}{3}}\varphi _{|m|-1}}\!;  \nonumber \\
&\varphi_{n} \!=\!  C_n \left( \frac{2\pi}{\sqrt{3} a^2} \right)^{1/4}  e^{-\frac12z_{k_2}^2(1+\frac{i }{\sqrt{3}})} \mathbb{H}_{n}(\!z_{k_2}\!),    \\
& C_n=\sqrt{\frac{3}{\sqrt{\pi}2^{(n+1)}n!}}, \qquad  z_{k_2}=\frac{3^{1/4}}{\sqrt{2\pi}}(\frac{\pi}{a}x_1 + k_2 a). \nonumber 
\end{align}
Here, 
$\mathbb{H}_m$ are Hermite polynomials, and the sign of $m$ identifies the conduction ($m>0$) and valence ($m<0$) band states.

The states in the basis set $|_{t}^{m}(\vect k)\rangle $ transform according to the irreducible representations of the symmetry group ${\cal M}_6^2$ of the vortex lattice field, which includes C$_6$ rotations and magnetic translations $ {\cal G} = \{\hat{\Theta}_{\vect X}
=e^{i \pi n_1' x_2/a} \hat{T}_{\vect X}, \vect X=n_1'a \hat{\vect{x}}_1+n_2'a \hat{\vect{x}}_2\} \subset {\cal M}_6^2$. 
In contrast to usual translations, magnetic translations do not commute with each other, $\Theta_{\vect{a}_1}\Theta_{\vect{a}_2}=-\Theta_{\vect{a}_2}\Theta_{\vect{a}_1}$, however, the group ${\cal M}_6^2$ contains 
an abelian subgroup,
\begin{align*}
{\cal G}' =\{\hat{\Theta}_{\vect R}
=e^{i \pi n_1' x_2/a} \hat{T}_{\vect R}, \vect R=2 n_1' a \hat{\vect{x}}_1+ 2 n_2' a \hat{\vect{x}}_2\},
\end{align*}  
which is formed by translations on a superlattice with isotropically doubled period and a unit cell area $4{\cal S}$. Because of this, it is possible to classify the states on magentic superlattice using the wave vector $\vect q$ taken over the folded Brillouin zone (BZ) with the area $\frac{\sqrt{3}}{8} g^2$, four times smaller than the BZ area of the gometrical vortex lattice. Each of these folded states is two-fold degenerate \cite{Zak,chen_prb_2014}, which is prescribed by the anti-commutation, $\Theta_{\vect{a}_1}\Theta_{\vect{a}_2}=-\Theta_{\vect{a}_2}\Theta_{\vect{a}_1}$, of the operators of elementary translations.  
By analysing characters of the magnetic translation group ${\cal M}_6^2$, one can find that the latter features six different 2-dimensional irreducible representations related to the states with the wave vector $\vect q =0$ in the centre of magnetic BZ. In practice, these six types of irredicible representations can be constructed using linear combinations of Bloch functions $|_{t}^{\pm (6M+N)}(\vect k)\rangle $ with different $M\ge 0$ but fixed $N=0,1,2,3,4,5$. 

Using the basis of Bloch functions in Eq.~\eqref{eq:bloch_wf},
Hamiltonian Eq.~\eqref{eq:MainHofPeriodicVSL}, 
can be represented in the form of the Heisenberg matrix, 
\begin{widetext}
\begin{align}
\label{eq:EntryofHeisenberg}
&
\langle_{t}^{m}(\vect k)| {\cal H} |_{\tilde t}^{\tilde m} (\tilde{\vect k} )\rangle  
= \frac{\hbar v}{a}\sqrt{\frac{4\pi}{\sqrt{3}}} \delta_{t,\tilde t} \delta_{\vect k ,\tilde{\vect k}} \left[\frac{m}{|m|} \sqrt{|m|}  \delta_{m,\tilde{m}} 
 + 
\sum_{n_1 n_2}  
\frac{e^{-i 2 n_1 k_1 a }}{\sqrt{(1+\delta_{m0})(1+\delta_{\tilde{m}0})}} 
 \frac{ \mu _{m, \tilde m, n_1,n_2}^{k_2 + \frac{\sqrt{3}}{4}gt, k_2 + \frac{\sqrt{3}}{4}g(n_1+t)} }{\omega_{n_1n_2}^2(1+\omega_{n_1n_2}^2 \alpha^{\text{-}2})} \right], \\
& 
\mu_{m, \tilde m,n_1,n_2}^{\kappa, \tilde \kappa} = \frac{m}{|m|}(n_1e^{i\pi/3}+n_2){\cal V}_{|m|-1,|\tilde m|}(n_2, \kappa, \tilde \kappa) 
- \frac{\tilde m}{|\tilde m|}(n_1e^{-i\pi/3}+n_2){\cal V}_{|m|,|\tilde m|-1} (n_2, \kappa, \tilde \kappa) \nonumber \\
&
{\cal V}_{N,\tilde N}(n_2, \kappa, \tilde \kappa) = \frac{1}{\sqrt{6}}  C_N C_{\tilde N} \int_{-\infty}^{\infty} \frac{dx_1}{a} 
\mathbb{H}_N (z_\kappa) 
\mathbb{H}_{\tilde N} (z_{\tilde \kappa})
e^{-\frac12 z_{\kappa}^2(1-\frac{i }{\sqrt{3}})} e^{-\frac12 z_{\tilde \kappa}^2(1+\frac{i }{\sqrt{3}})} e^{i\frac{\sqrt{3}}{2} n_2x_1g}
\nonumber
\end{align} 
\end{widetext}
where $z_{\kappa}=\frac{3^{1/4}}{\sqrt{2\pi}}(\frac{\pi}{a}x_1 + \kappa a)$.

In Figure \ref{fig:VortexSLBandWidth}, we show bands (shaded) and gaps (white intervals) in the spectrum obtained by numerical diagonalisation of Heisenberg matrix (\ref{eq:EntryofHeisenberg}). In this calculation we used 80 LLs to guaratee the convergence of the energies in the lowest 20 bands on the conduction and valence band sides, and we included all points $\vect{g}_{n_1n_2}$ in the reciprocal space such that $|\vect{g}_{n_1n_2}|\le 6 4 \pi/(\sqrt{3} a )$. The calculated energies are scaled with $v\sqrt{2\hbar e B} \equiv \frac{\sqrt{4\pi}}{3^{1/4}}\hbar v /a$ (energy of $|m|=1$ LL an homogeneous field $B$), and they are shown for various values of $4\pi \lambda^2 B/\phi_0$, parameter we use to characterise magnetic field distribution across the unit cell. The band diagram in Fig.~\ref{fig:VortexSLBandWidth} is electron-hole symmetric, and the hierarchy of bands and gaps in this spectrum is universal, as it can be applied to Dirac electrons in the field of a vortex lattice in a film of any isotropic type-II superconductor. To mention, due to the large demagnetisation factor of a film, vortices enter the film at the field much lower than the nominal first critical field of a bulk superconductor, hence, justify the regime of $4\pi \lambda^2 B/\phi_0 \ll 1$. We have set the upper side of the interval of $4\pi \lambda^2 B/\phi_0 \le 2$ shown in Fig.~\ref{fig:VortexSLBandWidth} at the lowest possible limit for the second critical field $H_{c2}$, knowing that at higher values of $4\pi \lambda^2 B/\phi_0$ the minibands would converge further towards the LL spectrum in a homogenous magnetic field.

The spectrum in Fig.~\ref{fig:VortexSLBandWidth} features a dispersionless zero-energy band that exactly coincides with $m=0$ LL in graphene, both by its energy, large gaps separating it from the rest of the spectrum, and capacity to accommodate electrons. 
Also, the first magnetic minibands, both on conduction and valence band side, are only slightly narrowed and follow almost exactly the energy of $m=\pm1$ LL. 
All the other magnetic minibands progressively broaden upon the increase of parameter $\alpha$, and some touch at certain values of $\alpha$. Such band  degeneracies occur due to occasional non-avoided crossings of $\vect q =0$ energy levels at the mini-BZ centre ($\gamma$-point). These occasional crossings are allowed by symmetry, because the energy levels $\epsilon_{m}^\gamma$ at the mini-BZ centre belong to one of six different irreducible representations, so that the closest levels do not necessarily 'repel' each other on the energy axis.

As different irredicible representations are built by mixing parent LL states which are, at least, $|\delta m| = 6$ apart, such crossings can be identified by analysing only the diagonal entries in the Heisenberg matrix (\ref{eq:EntryofHeisenberg}),  
\begin{widetext}
\begin{equation}
\epsilon_{m}^\gamma 
= \frac{m}{|m|} \frac{\sqrt{4\pi}}{3^{1/4}} \frac{\hbar v}{a} \sqrt{|m|} 
 \left[1 - \frac12 \sum_{n_1,n_2} 
\frac{e^{-\frac12 t_{n_1,n_2}}}
{(1 + \alpha^{-2}\omega^2_{n_1n_2})}\ _1\mathbb{F}_1 (1-|m|, 2, t_{n_1,n_2})
\right], 
\quad 
t_{n_1,n_2}=\frac{4 \pi \omega^2_{n_1n_2}}{\sqrt{3}},\nonumber
\label{GammaPoint}
\end{equation}
\end{widetext}
where $_1\mathbb{F}_1$ is Kummer's confluent hypergeometric function. Hence, we estimate that, within the interval $\alpha \le 2$ such crossings whould appear at $\alpha_{2,3}=0.98$  for bands 2 and 3, at $\alpha_{6,7}=0.90$ for bands 6 and 7, and $\alpha_{14,15}=1.03$ for bands 14 and 15, and these values are close to the band touching points found in the exact diagonalisation of Eq.~(\ref{eq:EntryofHeisenberg}). At a finite ${\bf q}$, such separation of the spectrum into six independent groups is no more possible.

The separation {\it vs} mixing of the groups of levels can be followed for the states at $q\ll \pi/a$ near $\gamma$-point, where off-diagonal matrix elements between the closest energy states can evaluated analytically \cite{footnoteLaguerre}, leading to a 2x2 matrix, 
\begin{align}
&
{\cal H}_{m+1,m} = \left(\begin{array}{cc}
\epsilon_{m+1}^\gamma & \!\!\!\!\! \!\!\!\!\! \qquad \tilde{v} \tau_m (q_x + iq_y \frac{m}{|m|}) \\
 \tilde{v} \tau_m^* (q_x - iq_y \frac{m}{|m|})  \!\!\!\!\! \!\!\!\!\! & \qquad \epsilon_{m}^\gamma
\end{array}\right), 
\nonumber \\
&
\tilde{v}_m=\frac{v}{4}
\sum_{n_1n_2} \frac{e^{-\frac{1}{2}t_{n_1n_2}}}{1 + \alpha^{\text{-}2}\omega^2_{n_1n_2}}
[ L^0_{|m|+\frac{m}{2|m|}-\frac{1}{2}}(t_{n_1n_2}) \nonumber \\
&
\qquad 
-\frac{t_{n_1n_2}}{\sqrt{(|m|+\frac{m}{2|m|})^2-1/4}}L^2_{|m|+\frac{m}{2|m|}-\frac{3}{2}}(t_{n_1n_2}) ],
\nonumber 
\end{align}
where $\tau_m = e^{i \frac{\pi}{6} (3-\frac{m}{|m|})}$ and $L_n^{\alpha}\!(x)$ are Laguerre polynomials ($m \ne 0,\pm1$). 
This matrix catches the Dirac-type edges of touching bands at $\alpha = \alpha_{m,m+1}$, confirmed 
by the numerically calculated dispersions plotted over the entire magnetic BZ in the inset 
in Fig.~\ref{fig:VortexSLBandWidth}. Note that velocity of all these 'second generation' Dirac electrons 
appears to be almost the same, $\tilde{v}_{m,m+1}\approx 0.75 v$, and, due to the spin and valley degeneracy 
which has to be factored additionally to the two-fold degeneracy of states in magnetic minibands at $\phi=\frac12 \phi_0$, 
each of the calculated orbital states in the folded magnetic BZ is 8-fold degenerate. 

\begin{figure}[htbp]
\centering
\includegraphics[width=0.47 \textwidth]{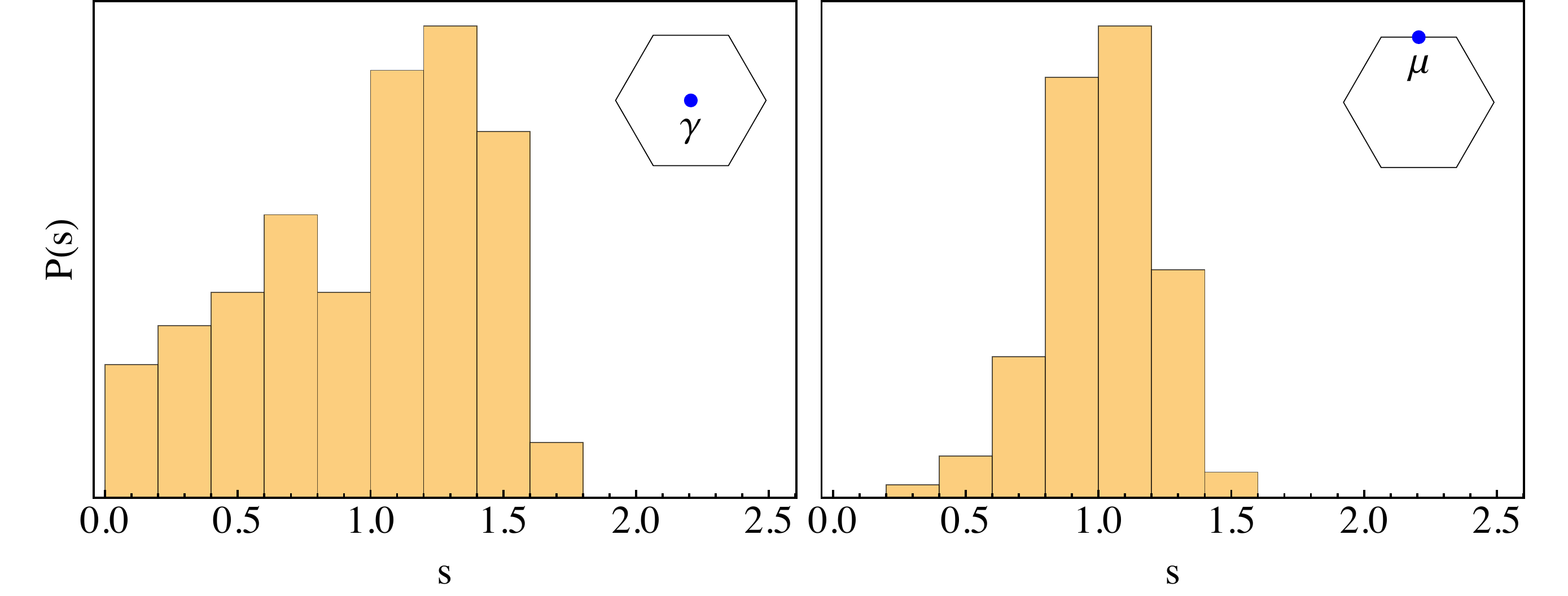} 
\caption{Distribution function $P(s)$ of the normalised level spacings in the miniband energies at $\gamma$ and $\mu$ 
points of the Broillouin minizone. Energy levels were taken from more than 30 states sampled ($E=0$ and the 1st band excluded) 
for 10 values of $0.5 \le \alpha \le 1.5$,  with a step of $\delta \alpha = 0.1$. 
}
\label{Pic2}
\end {figure}

The mixing {\it vs} separation  
of subsets of states in the magnetic minibands can also be traced using the distribution function $P(s)$ 
of the normalised level spacings, Fig.~\ref{Pic2} for two given points in the Brillouin minizone. 
For the $\mu$-point, it shows strong 'level repulsion' characteristic for the unitary symmetry class of random matrix theory \cite{Mehta}.
For the $\gamma$-point, levels can appear close to each other, as happens in other periodic systems \cite{Mucciolo,Silberbauer,Steffens,Dittrich} 
where high lattice symmetry splits the spectrum into subsets of states corresponding to different irreducible representations of the lattice symmetry group 
which can appear arbitrarily close to each other \cite{Fagas,Luna,Gumen,Mares,Sarkar}.

\begin{figure}[htbp]
\centering
\includegraphics[width=0.44 \textwidth]{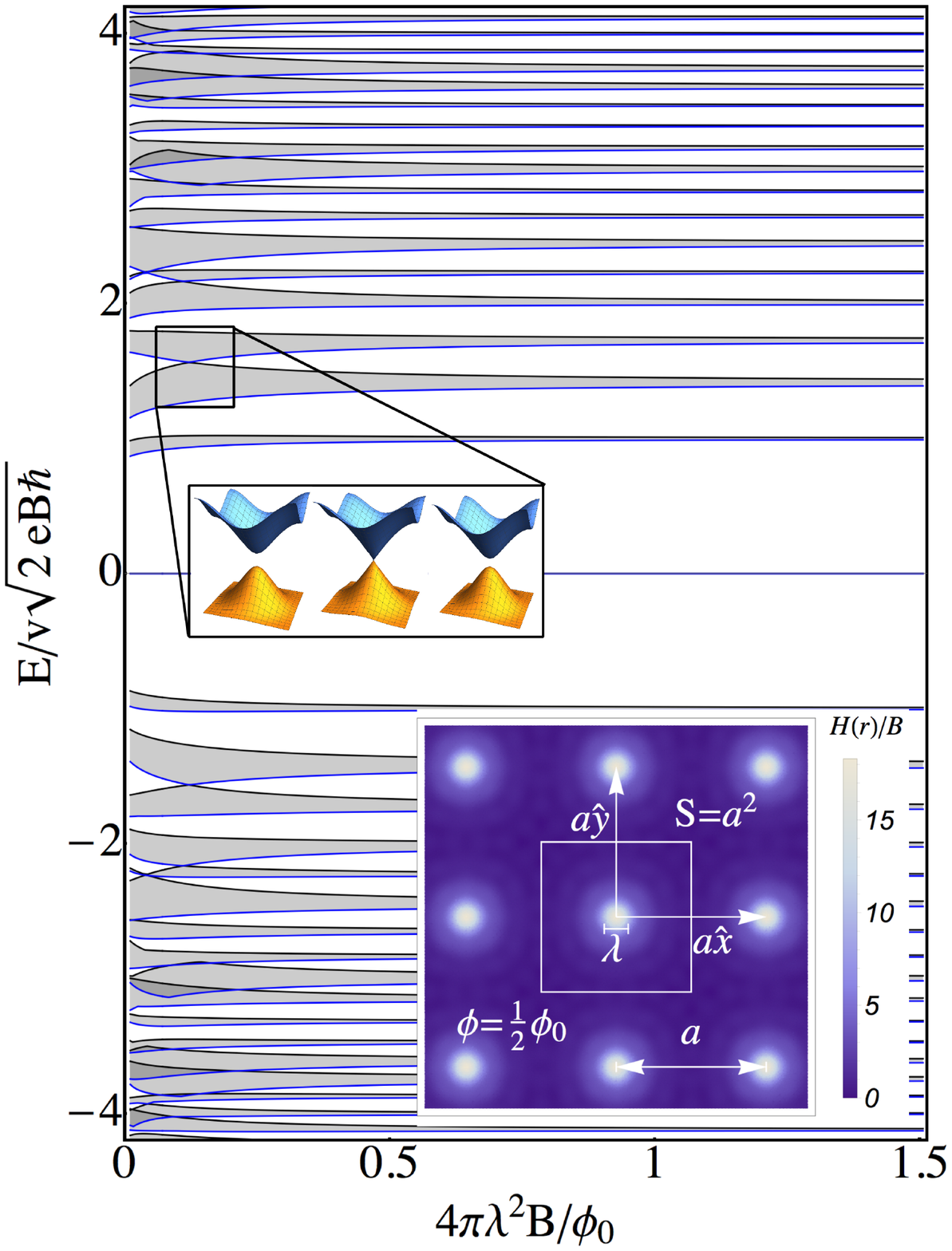} 
\caption{Spectrum of Dirac electrons in the presence of a square vortex lattice. 
Energy is scaled by $v\sqrt{2\hbar e B}$, and the spectrum includes one degenerate band at $E=0$.
Inset: 2nd and 3rd minibands dispersion over the folded magnetic Brillouin zone in the vicinity of their touching condition.}
\label{Pic3}
\end {figure}

The features we found for the Dirac electrons in hBN-encapsulated graphene placed over a hexagonal Abrikosov vortex lattice 
suggest that the quantum Hall effect at the filling factors $\nu = \pm 2$ and, to some extent, $\nu = \pm 6$ would remain a robust feature in the transport 
and capacitance measurements, whereas Shoubnikov - de Haas 
oscillations at higher filling factors would be strongly suppressed. This property of Dirac electrons in graphene seems to be generic for 
a broad range of magnetic field distributions. To stress this point, in Fig.~\ref{Pic3}, we show the calculated spetrum of  Dirac electrons moving in the a square vortex lattice, 
which has all the same characteristic features as the spectrum corresponding to the hexagonal vortex lattice.

{\it Acknowledgements.} We thank I.~Aleiner, A.K.~Geim, I.~Grigorieva and S.~Kubatkin for useful discussions. 
This work was funded by the European Graphene Flagship Project CNECT-ICT-604391, Royal Society, and ERC Synergy Grant Hetero2D.


\begin{thebibliography}{99}
\bibitem{FootnoteDirac}
In the non-ortogonal corrdinates we use here, the free-electron Dirac Hamiltonian in a magentic field has the form ($\lambda_{B}=\sqrt{\hbar/|eB_|}= a\sqrt{\sqrt{3}/(2\pi)}$)
$$\left(
\begin{matrix}
0  &  \hat{d}^{\dagger} \\
 \hat{d} &  0
\end{matrix}
\right),  
\hat{d} =\frac{-2}{\sqrt{3}}\left[ e^{ i\frac{ 2 \pi}{3}}\partial_{x_1} +  e^{ -i\frac{ 2 \pi}{3}}(\partial_{x_2}+  i \pi \frac{x_1}{a^2})\right]\!.\nonumber
$$
\bibitem{footnoteLaguerre}
We simplified Eq.~\eqref{eq:MainHofPeriodicVSL} using the identity \cite{Integral_Table},
\begin{align*}
&\int_{-\infty }^{\infty }e^{-x^2}\mathbb{H}_n(x+y)\mathbb{H}_m(x+z)dx\\
&=2^n\sqrt{\pi }m!y^{n-m}L_m^{n-m}(-2 y z), \ \ n\geq m,
\end{align*}
\end{thebibliography}
\end{document}